\documentclass[pre,aps,twocolumn,showpacs, showkeys,floatfix,nofootinbib,
superscriptaddress]{revtex4-1}
\usepackage[utf8]{inputenc}
\usepackage{amsmath}
\usepackage{amsfonts}
\usepackage{amssymb}
\usepackage{graphicx}

\usepackage[hidelinks,unicode=true]{hyperref}
\hypersetup{colorlinks=true,
	linkcolor=blue,
	urlcolor=blue,
	citecolor=blue,
	pdfhighlight=/N
}

\usepackage{comment}
\usepackage[normalem]{ulem}
\usepackage{ifpdf}
\usepackage{setspace}
\usepackage[dvipsnames]{xcolor}

\newcommand{\vaz}{\varnothing}
\newcommand{\AP}{\mathbf{A}}
\newcommand{\BP}{\mathbf{B}}
\newcommand{\ABP}{\mathbf{A\hspace{-0.18cm}B}}
\renewcommand{\vaz}{\times}
\renewcommand{\AP}{\circ}
\renewcommand{\BP}{\bullet}
\renewcommand{\ABP}{{\circ\hspace{-0.05cm}\bullet}}

\newcommand{\av}[1]{\langle {#1} \rangle}

\begin{document}

\title{Dynamical correlations and pairwise theory for the symbiotic contact process on networks}

\author{Marcelo M. de Oliveira}
\email{email: mmdeoliveira@ufsj.edu.br}
\affiliation{Departamento de Estatística, F\'{\i}sica e Matem\'atica,
	CAP, Universidade Federal de S\~ao Jo\~ao del Rei,
	36420-000 Ouro Branco, Minas Gerais - Brazil}

\author{Sidiney G. Alves}
\email{email: sidiney@ufsj.edu.br}
\affiliation{Departamento de Estatística, F\'{\i}sica e Matem\'atica,
	CAP, Universidade Federal de S\~ao Jo\~ao del Rei,
	36420-000 Ouro Branco, Minas Gerais - Brazil}

\author{Silvio C. Ferreira}
\email{email: silviojr@ufv.br}
\affiliation{Departamento de F\'{\i}sica, Universidade Federal de
	Vi\c{c}osa, 36570-900 Vi\c{c}osa, Minas Gerais, Brazil}
\affiliation{National Institute of Science and Technology for Complex
	Systems, Rio de Janeiro, Brazil}

\date{\today}

\begin{abstract}
The two-species symbiotic contact process (2SCP) is a stochastic process where
each vertex of a graph may be vacant or host at most one individual of each
species. Vertices with both species have a reduced death rate, representing a
symbiotic interaction, while the dynamics evolves according to the standard
(single species) contact process rules otherwise. We investigate the role of
dynamical correlations on the 2SCP  on homogeneous and heterogeneous networks
using pairwise mean-field theory. This approach is compared with the ordinary
one-site theory and stochastic simulations. We show that our approach
significantly outperforms the one-site theory. In particular, the stationary
state of the 2SCP model  on random regular networks is very accurately
reproduced by the pairwise mean-field,  even for relatively small values of
vertex degree, where expressive deviations of the standard mean-field are
observed. The pairwise approach is also able to capture the transition points
accurately for heterogeneous networks and provides rich phase diagrams with
transitions not predicted by the one-site method. Our theoretical results are
corroborated by extensive numerical simulations.
\end{abstract}


\maketitle

\section{Introduction}

Discontinuous transitions~\cite{DSouza2019}, observed in distinct classes of
cooperative systems, have lifted a renewed burst of interest within different
contexts such as social interactions~\cite{Perc2017,Gomez-Gardenes2016},
coinfections~\cite{Wang2019,Cai2015,Grassberger2016},
synchronization~\cite{Boccaletti2016,DSouza2019}, and
percolation~\cite{Boccaletti2016,DSouza2019}, to cite only a few fundamental
processes. A leading part of these investigations is addressed to phenomena
occurring on the top of complex networks~\cite{Albert2002}, that constitute
basic substrates for describing interacting patterns of complex
systems~\cite{PastorSatorras2015,Castellano2009,Ohtsuki2006}. While  in a
continuous transition the order parameter varies continuously from zero, in the
discontinuous case it suddenly jumps to a finite value  at the transition point,
involving a macroscopic portion of the system and this transition is commonly
refereed as catastrophic or abrupt~\cite{Strogatz2018}.

Coinfection epidemics~\cite{Wang2019}, when a host can be infected
simultaneously by two distinct diseases, can result in coexisting thresholds
when competitive interactions are considered~\cite{Newman2005,Newman2013}.
Nevertheless, cooperative or synergistic interactions result in richer phase
diagrams, which may include discontinuous phase transitions, and have been a
topic of intense research~\cite{Newman2013, Chen2013, Chen2013, Cai2015,
	Grassberger2016, Janssen2016, Cui2017,Liu}.  Cooperation and competition  have
been investigated on multi-species interacting models in distinct ecological
contexts~\cite{Tubay2013, Dobramysl2013, Court2013, Markham2013, Iwata2011,
	DeOliveira2012}. Particularly, symbiotic interactions were recently investigated
in a two-species contact process (2SCP)~\cite{DeOliveira2012}. In the standard,
single-species contact process (CP)~\cite{Harris}, individuals lay on the
vertices of a graph, which was originally a lattice, but it was later extended
to networks~\cite{Castellano2006,Ferreira2011}.  In the standard CP model, each
individual can reproduce asexually with rate $\lambda$, with its offspring
occupying one empty nearest-neighbor randomly chosen, or die with rate $\mu$
(which can be taken as $\mu=1$ without loss of generality). In the
2SCP~\cite{DeOliveira2012}, two species, A and B, inhabit the same graph. The
symbiotic interaction is modeled via a reduced death rate, $\mu< 1$, at sites
doubly occupied (by one individual of each species). With the exception of this
interaction, the two populations evolve independently according to the standard
CP.

Apart from its interest as an elementary model of symbiosis, the 2SCP is
fundamentally interesting for the study of nonequilibrium phase transitions.
Extinction represents an absorbing state, i.e., a frozen state without
fluctuations of populations~\cite{Marro2005}. On regular lattices, it was found
that the 2SCP exhibits a continuous phase transition in one and two
dimensions~\cite{DeOliveira2012}. However, the transition becomes discontinuous
in the regime of strong symbiosis if diffusion is introduced~\cite{DeOliveira2014}. 
The 2SCP was recently investigated in complete graphs
and random regular networks~\cite{SampaioFilho2018} and it was conjectured that
the nature of its transition changes at the upper critical dimension, from
continuous to discontinuous. The phase diagram determining the regions of the
2SCP' space parameter $\mu$ versus $\lambda$  was determined in the ordinary,
one-site mean-field level~\cite{DeOliveira2012, DeOliveira2014,	SampaioFilho2018}, 
which neglects all dynamical correlations with the assumption
of statistical independence among the states of individuals even if they are
nearest-neighbors. This assumption is a strong approximation that can make the
theory inaccurate even for high dimensional systems\footnote{For a regular
	lattice of dimension $d$ the typical distances scales as $\ell \sim N^{1/d}$.
	For these random graphs we have $\ell\sim \ln N$ corresponding to
	$d=\infty$~\cite{barabasi2016network}.} such as the complex
networks~\cite{Gleeson2011}, where the small-world property~\cite{Albert2002}
resembles a mean-field (fully connected) regime. Dynamical correlations can be
introduced using pairwise approximations~\cite{Marro2005,Ben-Avraham1992} where
the mean-field equations for pairs of connected vertices are considered. Even
still being inaccurate in low dimensions~\cite{Marro2005},  pairwise {correlations} highly
improve the theoretical results in the case of dynamical processes on  complex
{networks~\cite{Keeling1999,Eames2002,Gleeson2011,Gleeson2013,Mata2014}}. 

In the present work, we investigate the role of dynamical correlations in the
2SCP  using a homogeneous pairwise mean-field (PMF) theory. The theoretical
predictions are compared with stochastic simulations on different random
networks, including homogeneous, Poissonian, and scale-free degree
distributions~\cite{Barabasi1999}. We observe that the PMF theory substantially
improves the ordinary mean-field  in all investigated cases, being very accurate
for determining the transition points and phase diagrams. Moreover, the PMF
phase diagrams are more complex, showing discontinuous transitions with either
symbiosis parameter $\mu$ or rate infection $\lambda$ fixed in contrast with
the ordinary mean-field where only the former one can happen.

The remainder of this paper is organized as follows. In Sec.~\ref{sec:models} we
present the definitions of the model and networks used as substrates for 
simulations. The mean-field theories are developed in Sec.~\ref{sec:mean-field}
and Appendix~\ref{app:pair} while the simulation methods used in the present work
are described in Sec~\ref{sec:methods}. Section~\ref{sec:results} is devoted to
the comparison between mean-field theories and simulation outcomes. Finally, our
conclusions and prospects are drawn in Sec.~\ref{sec:conclu}.

\section{Model and networks}
\label{sec:models}

We investigate the symmetric 2SCP~\cite{DeOliveira2012}, in which the involved
rates are the same for both species. The model is defined for a networked
substrate as follows. Each vertex of the network  can hold at most one
individual of each species A and B.  So, the state $S$ of a vertex $i$ can be
empty, represented by $S_i=\vaz$, occupied by only one individual of type A 
($S_i=\AP$) or B ($S_i=\BP$), or occupied by one individual of each species,
represented by $S_i=\ABP$. The reproduction of new individuals of a given
species happens independently of each other according to the standard CP
rules~\cite{Marro2005,Harris}.  A vertex $i$ with one individual of type A,
$S_i=\AP$ or $S_i=\ABP$, creates an offspring of type A at a randomly selected
nearest-neighbor $j$ with rate $\lambda$ if $j$ does not have an individual of
type A (i.e., if $S_j=\vaz$ or $S_j=\BP$). Same rules and rates are used for the
replication of a type B. Any individual in a vertex $i$ spontaneously dies with
rate 1 if they are alone ($S_i=\AP$ or $\BP$) or with rate $\mu$, otherwise
($S_i=\ABP$). If $\mu=1$ the two processes evolve independently, but for $\mu <
1$ they interact {\em symbiotically} since the annihilation rates are reduced at
vertices where both species are present. The stationary 2SCP can evolve
asymptotically to a fully active state, where there is coexistence of both
species or to an inactive (absorbing) phase in which both species are extinct.
There are also two partly active states, where only one of A  or B species are
extinct that can be reached for specific initial conditions (the dynamics in
these partly active states is out of our interest since it is that of a standard
single-species CP).

In this work, we use random graph models as substrates on which the 2SCP takes
place. The  number of neighbors of a vertex $i$, the vertex degree, is denoted
by $k_i$. We consider both homogeneous and heterogeneous networks. In random
regular (RR) networks~\cite{Ferreira2013} all vertices have the same degree
$k_i=k$ and connections are performed at random following the configuration
model~\cite{Catanzaro2005} avoiding both multiple and self connections. In the
Erd\"os-Renyi (ER) model~\cite{Albert2002}, each pair of vertices is connected
with probability $p$. When the size of the graph $N\to\infty$, its degree
distribution is a Poissonian with a finite mean $\av{k}=pN$. Both RR and ER
models belong to the class of homogeneous degree networks, where large
deviations from the average value do not occur~\cite{barabasi2016network}, whereas
only RR has a strictly homogeneous  degree distribution. The third substrate is the
Barab\'asi-Albert (BA) model~\cite{Barabasi1999}, in which the networks are
generated through a preferential attachment mechanism~\cite{Barabasi1999}
starting from a fully connected set with $m_0+1$ vertices and vertices with $m$
edges being added one at a time. We used $m=m_0$ such that the average degree
$\av{k}=m$  can be fixed accordingly for comparison with the homogeneous case.
The BA degree distribution follows asymptotically a power law (PL), $p_k \sim
k^{-\gamma}$, with $\gamma=3$ and is representative of a highly heterogeneous
network possessing a heavy-tailed degree distribution.

\section{Mean-field theory}
\label{sec:mean-field}
We consider a homogeneous mean-field approximation where all vertices are
assumed to have the same degree $k_i=k$. In order to apply this theory to
heterogeneous cases of ER and BA we use the same equations of the homogeneous
theory  replacing $k$ by $\av{k}$. This strategy has been  used to compare the
transition points of CP obtained in a homogeneous PMF theory, given by
$\lambda_\text{c}=k/(k-1)$~\cite{Marro2005}, with the simulations on
heterogeneous networks using 
$\lambda_\text{c}=\av{k}/(\av{k}-1)$~\cite{Munoz2010,Ferreira2011}.

Let $[S]$ be the
probability that a vertex is in state $S$ and $[S,S']$ the joint probability
that a vertex is in the state $S$ and its nearest neighbor in  the state $S'$. 
Symmetries in the rates with respect to distinct species
imply that $[\AP]=[\BP]$ and $[S,S']=[S',S]$ for any pair of states $S$ and $S'$.
Also, if changing  all $\AP$ by $\BP$ 
does not alter the probabilities. 
So,   $[\AP,\AP]=[\BP,\BP]$, $[\AP,\BP]=[\BP,\AP]$, $[\AP,\ABP]=[\BP,\ABP] = 
[\ABP,\AP]=[\ABP,\BP]$, $[\vaz,\BP]=[\vaz,\AP] = [\BP,\vaz]=[\AP,\vaz]$.

Following this approach, one identifies three distinct one-site variables, $[\AP]$, $[\ABP]$, and $[\vaz]$, related by the closure relation
\begin{equation}
	[\ABP]+2[\AP]+ [\vaz]=1.
	\label{eq:clos_1site}
\end{equation}

The dynamical equations for these variables are readily derived as~
\begin{equation}
\frac{d[\AP]}{d t} = -[\AP]+\mu[\ABP]+\lambda([\vaz,\AP]+[\vaz,\ABP]-[\AP,\BP]-[\AP,\ABP]),
\label{eq:dAdt1}
\end{equation}
\begin{eqnarray}
\frac{d[\vaz]}{d t} & = & [\AP]+[\BP] -\lambda([\vaz,\AP]+[\vaz,\BP]+2[\vaz,\ABP])\nonumber\\
                    & = & 2[\AP]-2\lambda([\vaz,\AP]+[\vaz,\ABP]),
\end{eqnarray}                   
and
\begin{eqnarray}
\frac{d[\ABP]}{d t} & = &  -2\mu[\ABP]+\lambda([\BP,\AP]+[\AP,\BP]+[\BP,\ABP]+[\AP,\ABP])\nonumber\\ 
& =  &                   -2\mu[\ABP]+2\lambda([\AP,\BP]+[\AP,\ABP]).
\label{eq:dABdt1}
\end{eqnarray}

Equations~\eqref{eq:dAdt1} to \eqref{eq:dABdt1} are exact but not closed since
the evolution of the one-site probabilities depends on pairs. Cutting the
correlations at a vertex level with the approximation $[S,S']=[S][S']$, we
obtain the ordinary mean-field equations~\cite{DeOliveira2012}
\begin{equation}
\frac{d[\AP]}{d t} = -[\AP]+\mu[\ABP]+\lambda([\vaz]([\AP]+[\ABP])-[\AP]([\BP]+[\ABP])),
\label{eq:dAdt}
\end{equation}
\begin{eqnarray}
\frac{d[\vaz]}{d t} & = & 2 [\AP] -\lambda [\vaz]([\AP]+[\BP]+2[\ABP])
\end{eqnarray}                   
and
\begin{eqnarray}
\frac{d[\ABP]}{d t} & = &   -2\mu[\ABP]+2\lambda [\AP]([\BP]+[\ABP]).
\label{eq:dABdt}
\end{eqnarray}
Note that if one species is extinct, the above system reduces to the mean-field theory
for the one-species CP 
\begin{equation}
\frac{d[\AP]}{d t} = -[\AP]+\lambda(1-[\AP])[\AP]
\end{equation}
with a transition point at $\lambda = 1$. 

The stationary solution of Eqs.~\eqref{eq:dAdt} to \eqref{eq:dABdt}  is given by~\cite{DeOliveira2012}
\begin{equation}
[\AP] = [\BP]  = \frac{\mu  \left[ 2(1-\mu) - \lambda
	+ \sqrt{\lambda^2 - 4\mu (1-\mu)} \right] }{2 \lambda (1-\mu)}.
\label{pMFT}
\end{equation}
and
\begin{equation}
[\ABP]= \frac{\lambda [\AP] ^2}{\mu - \lambda [\AP] }.
\label{pabmft}
\end{equation}
For $\mu \geq 1/2$, $[\AP]$ grows continuously from zero at $\lambda=1$, marking
the latter value as the transition point. The activity grows linearly, $[\AP]
\simeq [\mu/(2\mu -1)](\lambda-1)$, in this regime. For $\mu < 1/2$, however,
the expression is already positive for $\lambda = \sqrt{4 \mu(1-\mu)} < 1$, and
there is a {discontinuous} transition at this point.

Further improvement in including dynamical correlations is given by a PMF
theory, in which the system is described in terms of pairwise variables
$[S,S']$. The dynamical equations of pairs will depend on triplets $[S,S',S'']$
in which a pair $S,S'$ of nearest-neighbors is connected to a vertex in the state $S''$ through  the
vertex in the state $S'$. Here, we neglect that $S$ and $S''$ can also be
connected, i.e., we assume the network does not form triangles having thus a
negligible clustering coefficient~\cite{Watts1998}, condition obeyed by all
networks considered in this work. Several symmetries such as
$[S,S',S'']=[S'',S',S]$ and $\AP\rightarrow\BP$  can be exploited to reduce the
number of independent triplets.

We now proceed with the standard pairwise approximation~\cite{Ben-Avraham1992,Mata2014}
\begin{equation}
[S,S',S'']=\frac{[S,S'][S',S'']}{[S']}
\end{equation}
and apply the generic closure relation
\begin{equation}
\sum_{S'} [S,S'] =[S]
\end{equation}
to obtain a set of seven dynamical equation for the independent pairwise
variables, given by Eqs.~\eqref{eq:dA_Bpar} to \eqref{eq:dAB_vazpar} in
Appendix~\ref{app:pair}. These equations, in addition to the one-site dynamical
equations~\eqref{eq:dAdt1} to \eqref{eq:dABdt1}, build a closed system. It is
worth mentioning that pairwise approximations have been presented for
one-dimensional chains and square lattices in Ref.~\cite{DeOliveira2014} while
our approach is valid for generic homogeneous graphs (conditioned to have a very
low clustering coefficient).

\section{Simulation scheme}
\label{sec:methods}
Our simulations of the 2SCP on networks were implemented using an optimized
Gillespie algorithm~\cite{Cota}, in which we maintain two lists, one of singly
and another of doubly occupied vertices. Let $N_1$ and $N_2$ denote,
respectively, the numbers of such nodes. The total rate of (attempted)
transitions is $\lambda N_1+2\lambda N_2 +N_1+2\mu N_2\equiv (\Delta t)^{-1}$,
where $\Delta t$ is the average time increment associated with a given step of
the simulation.

At each time step, we randomly choose  among the events: (i) creation attempt by
an isolated individual, with probability $\lambda N_1\Delta t$; (ii) creation
attempt by an individual at a doubly occupied node, with probability $2\lambda
N_2\Delta t$; (iii) death of an isolated individual, with probability $N_1\Delta
t$; (iv) death of an individual at a doubly occupied node, with probability
$2\mu N_2$.

Once the event type is selected, a vertex $i$ is chosen at random from the
respective list. Creation attempt occurs at a  vertex $j$ randomly selected
among the nearest-neighbors of $i$. If $j$ is already occupied by an individual
of the species to be created, no change of state is implemented, and the
simulation continues to the next step. If node $i$ is doubly occupied, the
species of its offspring in a creation event is chosen to be $A$ or $B$ with
equal probability. In the same way, in an annihilation event at a
doubly-occupied node, we choose the species to be removed at random. Time is incremented
by $\Delta t$.

In the simulations, we sample the quasistationary (QS) distribution of active
nodes employing the QS simulation method~\cite{DeOliveira2005}. This scheme
consists in replacing the absorbing state, every time the system attempts to
visit it, with an active configuration randomly taken from the history of the
simulation. This procedure optimizes the numerical simulations restricting the
dynamics of the process to active states, and is a powerful tool in analyzing
continuous absorbing state phase transitions in networks
\cite{Ferreira2011,Ferreira2016a,Mata2014,DeArruda2015a,Valdano2015}. For a
recent analysis on such simulation methods applied to networks, see
Ref.~\cite{Sander2016}. In this work, we performed QS simulations for systems of
sizes up to $N = 10^6$ nodes, with each run lasting at least $10^6$ time units.
Averages are taken in the QS regime, after discarding an initial transient which
depends on the system size and symbiosis strength used.

Finally,  mean-field analysis were obtained through numerical integration of
the respective dynamical systems using fourth order Runge-Kutta method with a
time step $\Delta t   = 10^{-5}$. The steady state is computed after a
relaxation of $t>10^6$.

\section{Results and discussion}
\label{sec:results}
\begin{table*}
	\centering
	\def\arraystretch{1.5}
	\begin{tabular*}{0.95\linewidth}{c @{\extracolsep{\fill}}||ccc||cccc||cccc}
		\hline \hline
		&   \multicolumn{3}{c||}{$\av{k}=3$}  & \multicolumn{4}{c||}{$\langle k\rangle=6$} & \multicolumn{4}{c}{$\langle k\rangle=10$} \\
		\hline  
		 ~ $\mu$  ~ &  PMF  & RR & ER & PMF  & RR& ER & BA & PMF  & RR & ER & BA  \\
		\hline
		0.1  & 0.785 & 0.795(5) & 0.780(5) & 0.6857   & 0.680(5)& 0.678(2) & 0.790(5)   & 0.6587   &0.647(5)& 0.645(5) &0.725(2)  \\
		0.2  & 1.055 & 1.075(5)& 1.065(5) & 0.9014   & 0.905(5)& 0.905(5)  & 1.030(5)  & 0.8567   &0.855(5) & 0.865(5) &0.950(2) \\
		0.5  & 1.445 & 1.535(5)& 1.440(5)& 1.1872   &1.200(5) & 1.190(5) & 1.217(1)  & 1.1060   &1.11(1) & 1.105(5) & 1.11(1) \\
		1.0  & 1.5 & 1.625(5) & 1.495(5) & 1.2     & 1.220(5)& 1.195(5)  &  1.221(1)   & 1.111     &1.12(1)& 1.13(1) & 1.13(1) \\
		\hline \hline 
	\end{tabular*} 
	\caption{\label{tab:threshold} Transition points  of the control parameter,
		$\lambda_\text{c}$ for different values of $\mu$ obtained from the PMF theory
		and simulations in RR,  ER and BA networks using  average degrees $\langle
		k\rangle=3$, $6$, and  $10$.}
\end{table*}

\begin{figure}[!hbt]
	\includegraphics[width=0.97\linewidth]{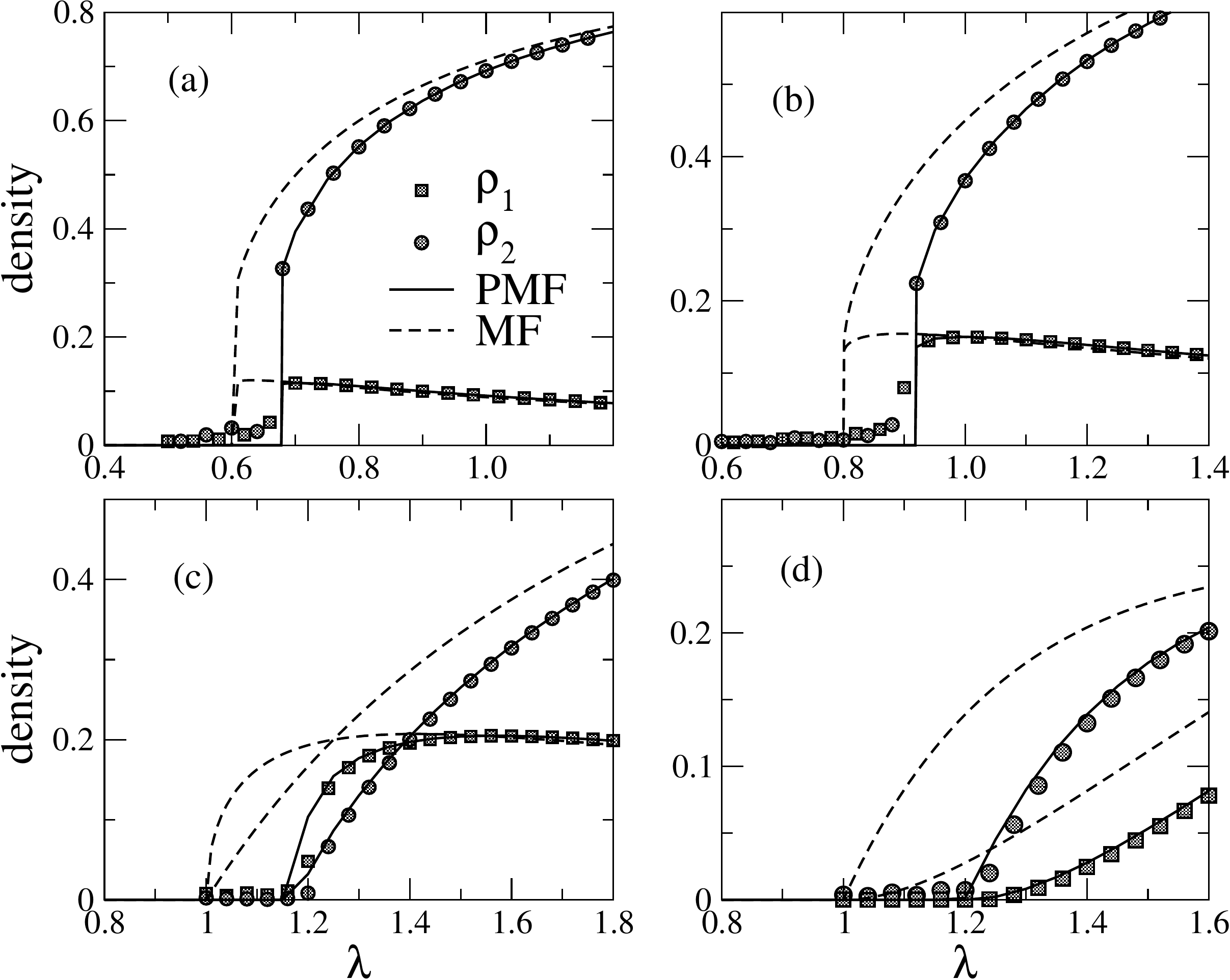}
	\caption{2SCP on RR networks of fixed degree $k=6$. The QS density of single
		(circles) and doubly (squares) occupied vertices are shown as functions of the
		infection rate for (a) $\mu=0.1$, (b) $\mu=0.2$, (c) $\mu=0.5$, and (d) $\mu=1$. The dashed curves represent the one-site while solid ones the
		pairwise mean-field theories. The system size used in simulations is $N=10^5$.}
	\label{RRq6}
\end{figure}

The quasistationary densities of sites occupied by a single
$\rho_1=\rho_\BP+\rho_\AP$ and by two species $\rho_2=\rho_{\ABP}$ are compared
with mean-field approximations in Fig.~\ref{RRq6} for a RR network with
connectivity degree $k=6$, using distinct values of the symbiotic strength
parameter $\mu$. This dynamics can exhibit bistability with both active and
absorbing stable stationary states~\cite{DeOliveira2012}. So, we first analyze 
an initial condition having all vertices in a doubly occupied state $\rho_2
=1$  such that this threshold represents the loss of global stability of the
absorbing state, marking the {\em lower spinodal} point.
We observe that the ordinary mean-field theory, although qualitatively
predicting the discontinuous or continuous nature of the transition, is not
quantitatively accurate  as indicated by the dashed curves in Fig.~\ref{RRq6}.
However, a far better result is obtained with the  PMF theory, represented by
the solid curves,  presenting  an excellent agreement between theory and
simulations, for both near and above the transition point regimes  irrespective
of $\mu$ values. Increasing the connectivity $k$ reduces the transition value of
$\lambda$ as can be seen in Table~\ref{tab:threshold}. Actually, the threshold
will converge to the one-site mean-field value for $k\gg 1$ since this limit
corresponds to the fully connected graph for which the one-site theory is
exact in the thermodynamical limit.

\begin{figure}[!hbt]
	\includegraphics[width=0.97\linewidth]{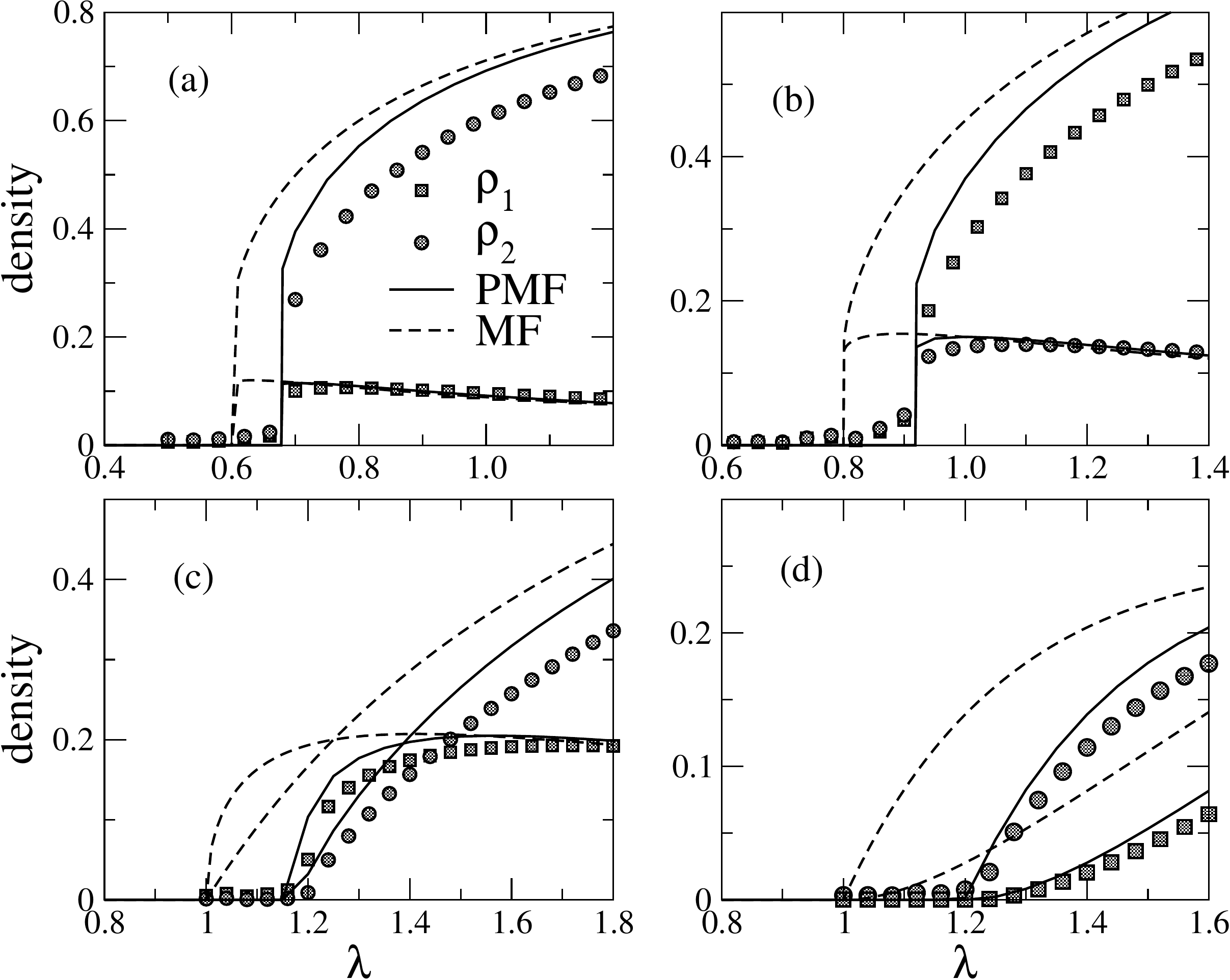}
	\caption{2SCP on ER networks with average degree  $\av{k}=6$. The QS density of
		single (circles) and doubly (squares) occupied vertices are shown as functions of
		the infection rate for  (a) $\mu=0.1$, (b) $\mu=0.2$, (c) $\mu=0.5$, and (d) $\mu=1$. The
		dashed curves represent the one-site while solid ones the pairwise mean-field
		theories. The system size used in simulations is $N=10^5$. }
	\label{ERq6}
\end{figure}
Figure~\ref{ERq6} compares the results from simulations and mean field theories
for ER networks with $\av{k}=6$ using the substitution of $k$ by the mean
connectivity $\langle k\rangle$. As observed in RR networks, the one-site
approximation reproduces qualitatively well the nature of transition but
underestimates the transition point. The PMF accurately matches the transition
point and density of singly occupied vertices in the active phase   of
simulations but underestimates the density of doubly occupied vertices and
consequently the overall density of active vertices $\rho=\rho_1+\rho_2$.

\begin{figure}[!hbt]
\includegraphics[width=0.97\linewidth]{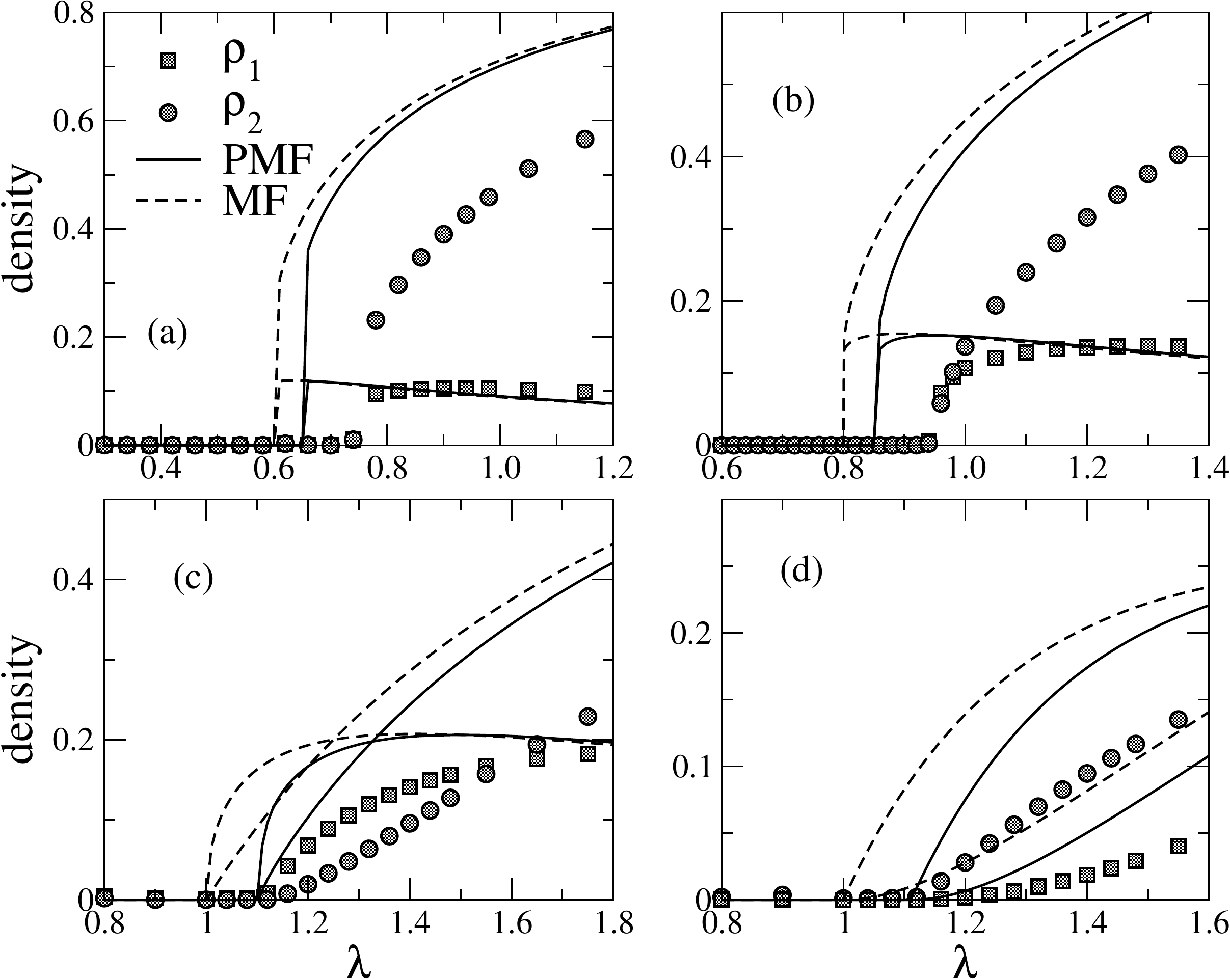}
\caption{2SCP on BA networks with average degree  $\av{k}=10$. The QS density of
	single (circles) and doubly (squares) occupied vertices are shown as functions
	of the infection rate for (a) $\mu=0.1$, (b) $\mu=0.2$, (c) $\mu=0.5$, and (d) $\mu=1$. The
	dashed curves represent the one-site while solid ones the pairwise mean-field
	theories. The system size used in simulations is $N=10^6$.}
\label{BAm5}
\end{figure}

The role of heterogeneity is further investigated in Fig.~\ref{BAm5} where  the
results on  BA networks with $\langle k\rangle=10$ are shown. The PMF theory
outperforms  the ordinary one but still underestimates the location of the phase
transition for small values of $\mu$, where the phase transition is
discontinuous while  yields a very good agreement for the transition point for
the continuous cases with $\mu>0.5$ despite of the high heterogeneity of the BA
networks, where hubs are expected to play some important role. This finding in
the continuous case is in agreement with one-species CP on scale-free
networks~\cite{Munoz2010,Ferreira2011}.

The transition points for the investigated networks (RR, ER and BA) with
different average degrees are presented in Table~\ref{tab:threshold}. We observe
that the PMF theory exhibits a higher accuracy for large average connectivity,
as expected, since in the limit $\av{k}\gg 1$ corresponds to the fully connected
graph where  mean-field theories become exact. Another important observation is
that the PMF performs worst for BA networks when $\mu$ is small while the
accuracy tends to be reduced for larger $\mu$ in RR networks. In BA networks, where
hubs are present, small $\mu$ will enhance activity localized in the
neighborhood of hubs which, in turn, are not sufficiently mixed for the regime
of scale-free networks with $\gamma=3$~\cite{Hoyst2005}. More precisely, the
star subgraph containing a hub and its nearest-neighbors can stay active in
isolation for long periods through a feedback mechanism where the hub activates
its neighbors that in turn reactivate the hub
recurrently~\cite{Chatterjee2009,Boguna2013,Ferreira2016a}. Since localization
opposes to the homogeneous mixing hypothesis of the mean-field methods, larger
deviations from the theory are indeed expected in this regime. Another
intriguing feature of the data of Table~\ref{tab:threshold} is that the PMF
thresholds are nearer to the simulations for the slightly heterogeneous ER  than
from the homogeneous RR networks, which becomes more evident for lower $\av{k}$
and higher $\mu$. A similar phenomenon was also observed  for the one-species CP
on networks~\cite{Mata2014} and associated with the approximation $k\approx
\av{k}$, which would not be observed in a degree-based mean-field
theory~\cite{PastorSatorras2015}.
 
\begin{figure}[tbh!]
	\centering
	\includegraphics[width=0.8\linewidth]{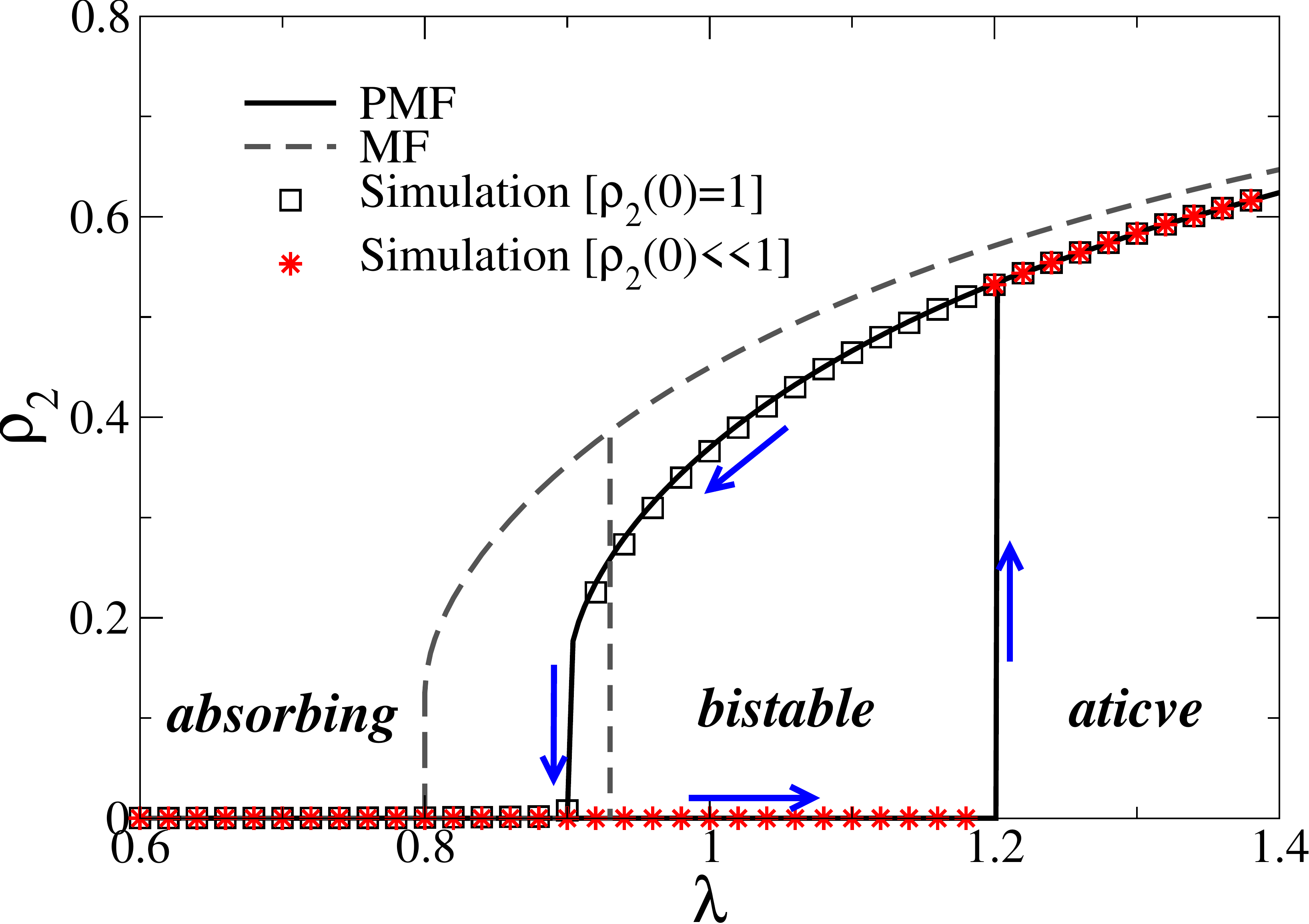}
	\caption{Hysteresis analysis for 2SCP on RR networks with $k=6$ and $\mu=0.2$.
		Solid curves: PMF theory (arrows indicate the hysteresis flow). Dashed curves:
		ordinary mean-field theory. Symbols: results from simulations for the RR
		network with $N=10^5$ vertices using different initial conditions.}
	\label{fig:hysterisis}
\end{figure}

Up to this point, we have addressed the global loss of the stability of the
absorbing state that, in the case of discontinuous transitions, involves  the
transition from an absorbing to a bistable stationary state where both absorbing
phase or active  phases can be stable depending on the initial condition. We now
also consider the loss of global stability of the active phase, i. e., the
transition from bistable to active phases. For this aim, we  consider an initial
condition very close to the absorbing state with a very low density of doubly
occupied vertices ($\rho_1=0$ and $\rho_\ABP \ll 1$)  in complement to the fully
occupied state $\rho_\ABP=1$ used previously. We can, therefore, compute
hysteresis curves where the phase coexistence can be derived as  in
Fig.~\ref{fig:hysterisis} where one sees absorbing, bistable and active regions.

\begin{figure}[hbt]
	\includegraphics[width=0.8\linewidth]{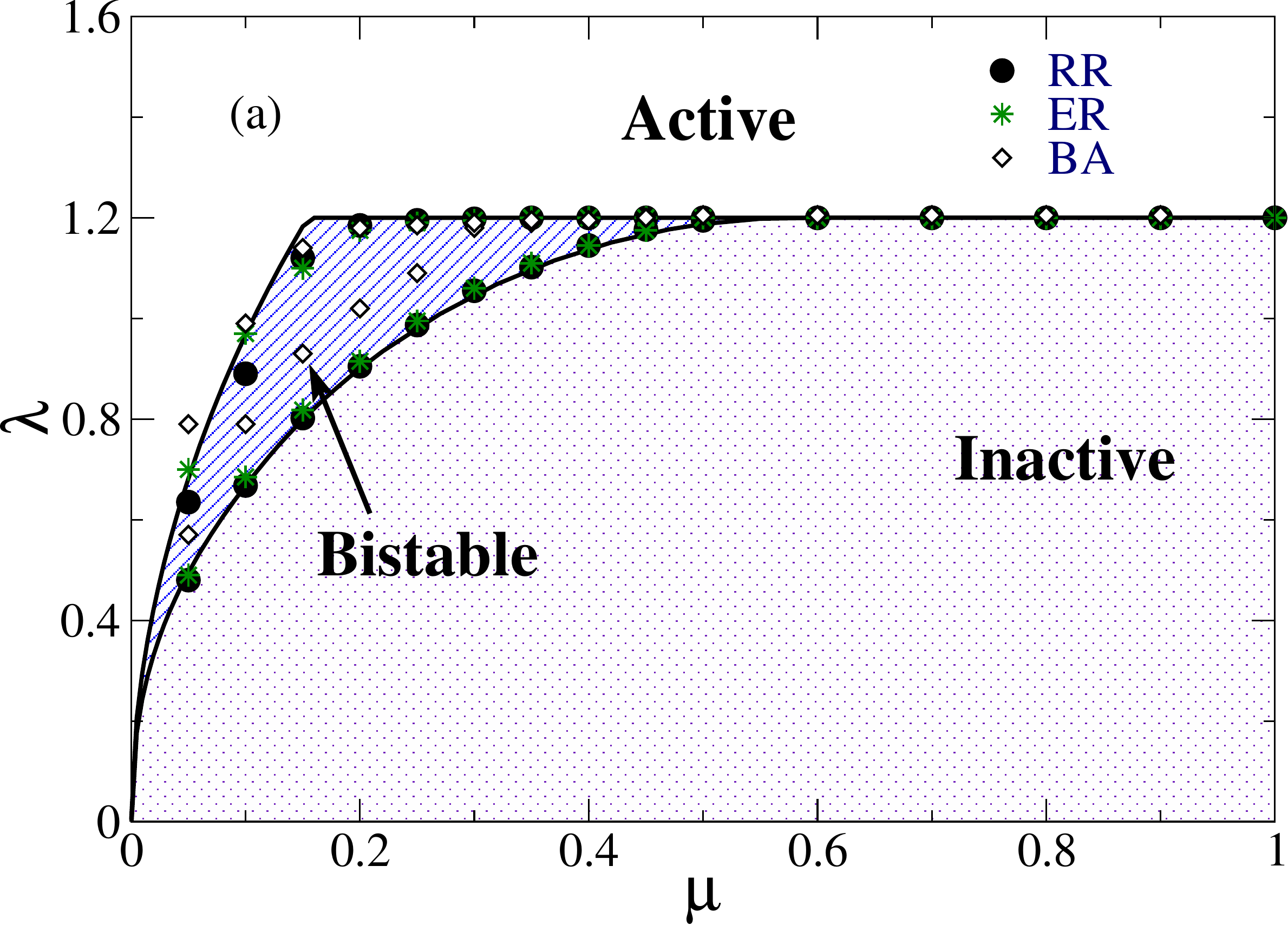}\\
	\includegraphics[width=0.8\linewidth]{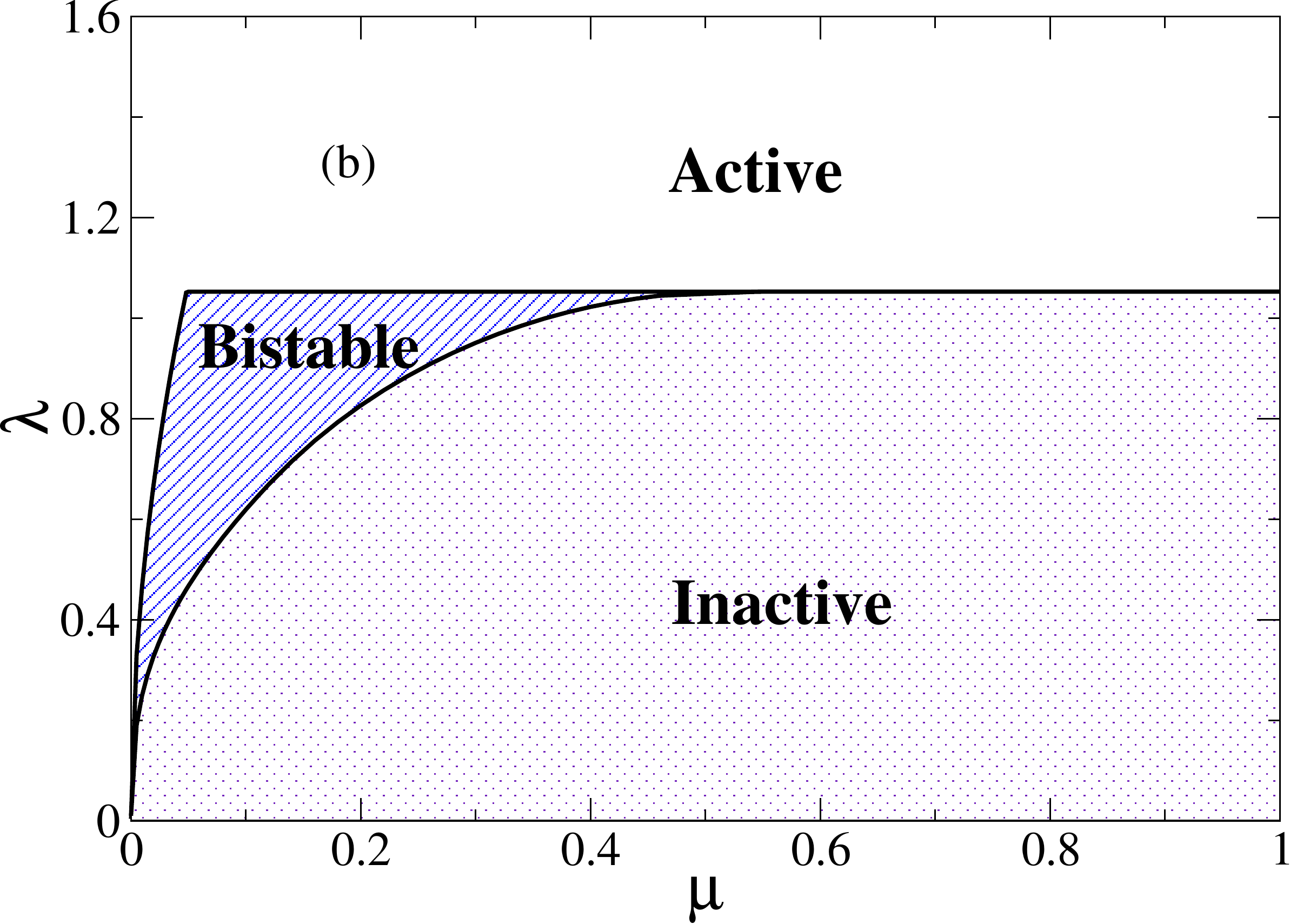}\\
	\includegraphics[width=0.8\linewidth]{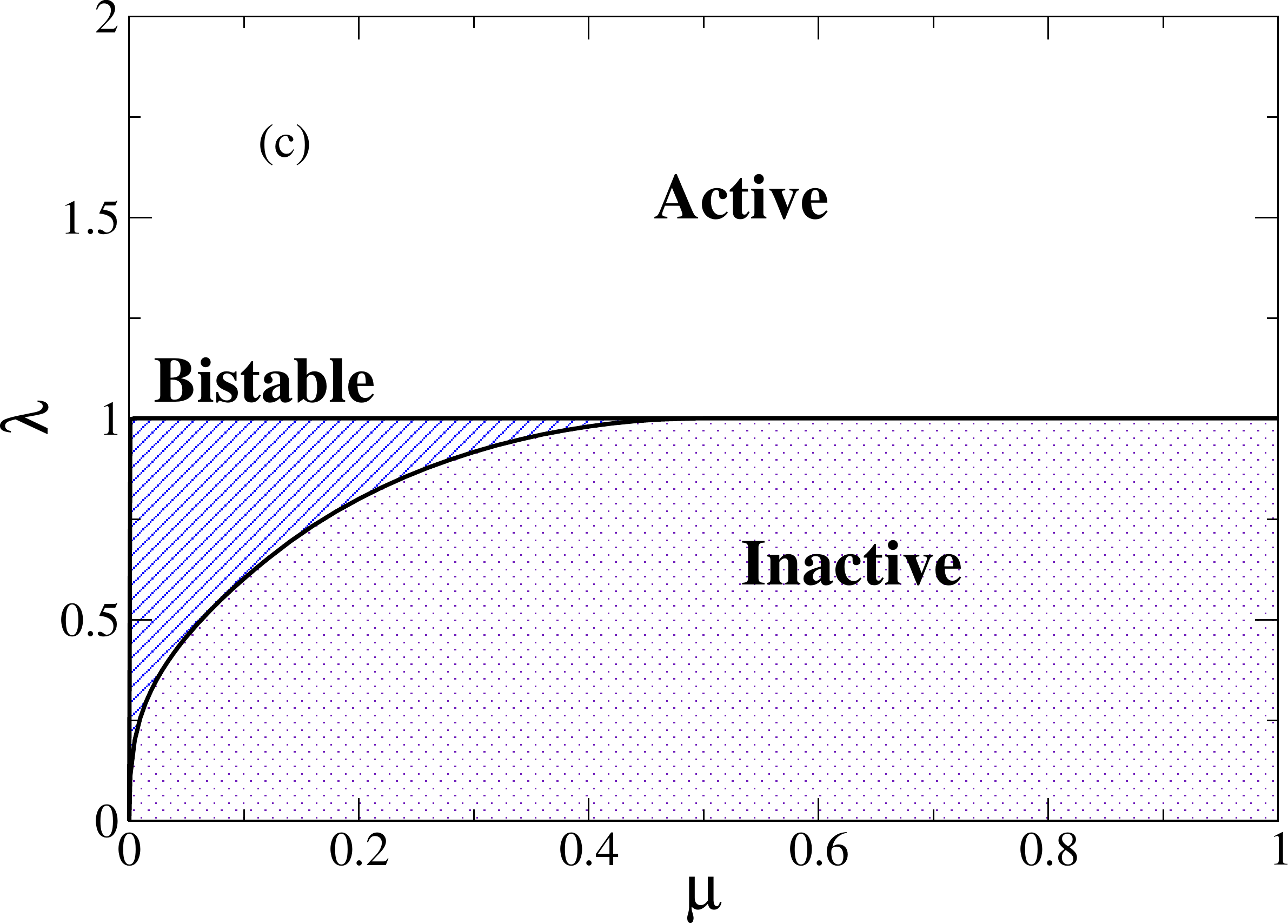}
	\caption{Phase diagram for the 2SCP obtained from PMF theory for
		distinct average degree (a) $\av{k}=6$, (b) $\av{k}=20$, and (d)
		$\av{k}=10^3$. Hashed and dotted regions represent bistable and inactive
		phases respectively while the empty region corresponds to the active phase of
		PMF theory. Symbols are the results extracted from numerical
		simulations of different networks models with $N=10^5$. }
	\label{phased}
\end{figure}
Phase diagrams in the parameter space  $(\lambda,\mu)$ obtained via PMF for
different connectivities are shown in Fig.~\ref{phased}. The diagrams exhibit
three connected regions corresponding to globally inactive ({dotted}
region), globally active (empty region) and bistable (hashed region).  An
interesting difference with respect to the ordinary mean-field is a transition
from a globally stable active phase to a bistable dynamics obtained either by
fixing the infection rate $\lambda$ and varying the symbiotic parameter $\mu$ or
using fixed $\mu$ and changing $\lambda$ whereas in an ordinary mean-field
theory such a transition occurs only for fixed $\mu$; see, for example, Fig. 3
of Ref.~\cite{SampaioFilho2018} which resembles very much the  limit of large
degree shown in Fig.~\ref{phased}(c). Indeed, the region with double transitions
(inactive to bistable and bistable to active) at fixed $\lambda$, occurring in
the small $\mu$ region, shrinks as the connectivity increases towards a fully
connected graph limit as shown in Figs.~\ref{phased}(b) and (c). {Triple
	points  shown in  the phase diagrams  depend on connectivity, departing from
	$(\lambda_\text{t},\mu_\text{t})=(1,1/2)$ for the complete graph (one-site
	mean-field)~\cite{DeOliveira2012} to $\lambda_\text{t}=\tfrac{\av{k}}{\av{k}-1}$
	and $\mu_\text{t}=0.501(1),~0.545(5)$, and $0.595(5)$ for $\av{k}=10^3$, $20$, and
	$6$, respectively.}
 {It worths to comment that phase diagrams with qualitatively similar
 	spinoidals found in 2SCP with our PMF theory have been reported for other
 	models with abrupt transitions on networks~\cite{Majdandzic2014,
 		Bottcher2017a, Bottcher2017a, Valdez2016} with a difference that they were
 	obtained by a first-order  mean-field theory, whereas in the 2SCP
 	model we needed to go up to a second-order pairwise theory.}

Figure~\ref{phased}(a) also presents the phase diagram obtained via simulations
on different graphs with average degree $\av{k}=6$. One observes a remarkable
good match between simulations and theory for RR and ER (homogeneous) networks.
The quantitative agreement is less satisfactory for BA model, but it is still
qualitatively correct. The dynamical correlations in these infinite dimensional
systems have proven to lead not just to a quantitatively better theory but also
to new behaviors not predicted by the ordinary mean-field theory.

\section{Conclusions}
\label{sec:conclu}

Stochastic interacting systems with cooperative or synergistic couplings can be
used for modeling different phenomena, such as coinfections of pathological
agents, social relations, and ecological interactions, among others. One key
issue of these models involves the nonequilibrium phase transitions among active
and absorbing phases that can emerge in some systems, especially the abrupt
ones, where macroscopic order parameters change discontinuously. Moreover,
understanding the behavior of these transitions taking place on the top of
complex networks  has become very relevant since most of these
dynamical processes indeed involve interactions mediated by networked systems.
In the present work we investigate a simple model, in which  two species lying
on  networks interact symbiotically, the symbiotic contact
process~\cite{DeOliveira2012}. We used both stochastic simulations and
mean-field pairwise approximations, the latter reckoning dynamical correlations
not considered in previous analytical (one-site mean-field) studies of this
model.

The PMF theory outperforms the ordinary one and, more importantly,
predicts features  not captured by the latter, in agreement with the simulations
on distinct types of networks with homogeneous, Poissonian, and scale-free
degree distributions. In the case of homogeneous networks, the PMF theory is
accurate describing very well both the dynamics near and above the transitions
(supercritical region). Despite of disregarding the heterogeneity of the
networks, our theory also gives very good estimates of the transition points  in
the case of Poissonian and scale-free networks but cannot quantitatively capture
the prevalence in the highly active regime.

We also investigated the phase diagrams exhibiting globally absorbing, bistable,
and globally active regions. The PMF theory, beyond being quantitatively more
accurate than the ordinary one, provides a much richer phase diagram where
discontinuous transitions, involving bistable phases, can be obtained either
varying the infection rate $\lambda$ or the symbiosis parameter $\mu$
independently, whereas this transition happens only at fixed $\mu$ in the
one-site approach. These PMF results are confirmed by  extensive numerical
simulations. Our findings provide an important example of the role played by
dynamical correlations in cooperative processes on networked substrates, which
is commonly not considered in other theoretical approaches.
{Pair-approximations as the one developed in this work can be used to
	improve the accuracy of phase diagrams   observed in other related
	works~\cite{Majdandzic2014,Bottcher2017} where first order mean-field theory
	captures the qualitative behavior  but analytical spinoidals deviate from
	simulations when the average connectivity is reduced.} Finally, we expect that our work
will stimulate future analysis of dynamical correlations
on other cooperative processes.

The theories presented in this paper do not include the network heterogeneity
and, specially, the degree distribution. As a prospect for forthcoming analysis,
one could use degree-based~\cite{PastorSatorras2015} or
individual-based~\cite{VanMieghem2012} mean-field theories to tackle the role
of heterogeneity.

\begin{acknowledgments}
The authors thank the financial support of Conselho Nacional de Desenvolvimento
Cient\'{i}fico e Tecnol\'{o}gico (CNPq) and Funda\c{c}\~{a}o de Amparo \`a Pesquisa do
Estado de Minas Gerais (FAPEMIG).
\end{acknowledgments}

\appendix
\begin{widetext}

\section{Pair mean-field theory}
\label{app:pair}

In the pair approximation, the  seven two-site variables to be considered are
$[\AP,\BP]$, $[\AP,\AP]$ $[\vaz,\AP]$, $[\AP,\ABP]$, $[\ABP,\ABP]$,
$[\vaz,\ABP]$, and $[\vaz,\vaz]$. All other combinations are equivalent to some
of them, due to symmetry. Considering that every vertex has $k$
nearest-neighbors, after some algebra we obtain the following set of dynamical
equations
\begin{eqnarray}
\frac{d [\AP,\BP]}{dt}  & = &  -2(1+{\lambda}/{k})[\AP,\BP]+2\mu [\AP,\ABP] + 2\lambda \frac{k-1}{k} 
          ([\AP,\vaz,\BP]+[\AP,\vaz,\ABP]-[\AP,\BP,\AP]-[\AP,\BP,\ABP])\label{eq:dA_B} 
\end{eqnarray}
\begin{eqnarray}
\frac{d [\AP,\ABP]}{dt} & = &   -(1+2\mu+\lambda/k)[\AP,\ABP]+ \mu [\ABP,\ABP]  + \frac{\lambda}{k}([\AP,\BP]+[\vaz,\ABP])+\nonumber \\  
& &  + \lambda \frac{k-1}{k}([\AP,\vaz,\ABP] + [\ABP,\vaz,\ABP]+ 
        [\AP,\AP,\BP]+[\AP,\AP,\ABP]
        + [\AP,\BP,\AP] -[\ABP,\AP,\ABP]). \label{eq:dA_AB}
\end{eqnarray}

\begin{eqnarray}
\frac{d[\vaz,\vaz]}{dt}   & = & 4[\AP,\vaz]-4\lambda \frac{k-1}{k}([\vaz,\vaz,\AP]+[\vaz,\vaz,\ABP])\label{eq:dvaz_vaz}
\end{eqnarray}

\begin{eqnarray}
\frac{d[\AP,\AP]}{dt} & = &  2(-[\AP,\AP]+\mu[\AP,\ABP] +\frac{\lambda}{k}[\AP,\vaz]) +
  2\lambda \frac{k-1}{k}([\AP,\vaz,\AP]+[\AP,\vaz,\ABP]-[\AP,\AP,\BP]-[\AP,\AP,\ABP]).\label{eq:dA_A}
\end{eqnarray}

\begin{eqnarray}
\frac{d [\ABP,\ABP]}{dt} & = & -4[\ABP,\ABP] +4\frac{\lambda}{k}[\AP,\ABP]+4\lambda \frac{k-1}{k} (
[\ABP,\AP,\ABP]+[\ABP,\AP,\BP]) \label{eq:dAB_AB}
\end{eqnarray}

\begin{eqnarray}
\frac{d [\AP,\vaz]}{dt} & = & -[\AP,\vaz] -\frac{\lambda}{k}[\AP,\vaz] + 
\mu [\ABP,\vaz]+[\AP,\AP]+[\AP,\BP]+\nonumber \\
& & + \lambda \frac{k-1}{k}([\AP,{\vaz,\vaz}]+[\ABP,{\vaz,\vaz}]- 
[{\AP,\vaz},\AP]-2[{\AP,\vaz},\ABP]- [{\AP,\vaz},\BP]
-[\BP,{\AP,\vaz}]-[\ABP,{\AP,\vaz}]). \label{eq:dA_vaz}
\end{eqnarray}

\begin{eqnarray}
 \frac{d[\ABP,\vaz]}{dt}  & = & -2(\mu+\lambda/k)[\ABP,\vaz]+2[\ABP,\AP]+2\lambda\frac{k-1}{k}([\AP,\BP,\vaz]+[\ABP,{\AP,\vaz}] -[\ABP,\vaz,\AP]-[\ABP,\vaz,\ABP]).\label{eq:dAB_vaz}
\end{eqnarray}

Let us explain in details the terms in Eq.~\eqref{eq:dA_B}. The derivation of
the remaining equations follows from a similar reasoning. Thus, in
Eq.~\eqref{eq:dA_B}, the term proportional to $[\AP,\BP]$ represents two
contributions. The one proportional to $-2$ is the spontaneous annihilation of
each individual ($\AP\to \vaz$ and $\BP\to \vaz$). The term proportional to
$2\lambda/k$ represents the creation of each individual of this pair into the
other ($\AP,\BP\to  \ABP,\BP$ and $\AP,\BP\to  \AP,\ABP$). The term proportional
to $[\AP,\ABP]$ is due to the annihilation with rate $\mu$ of each individual in
the doubly occupied vertex  ($\ABP\to\AP$ and $\ABP\to\BP$). Finally, the triplet
terms represent the creation of outsiders into the pair. For example, the first
triplet term represents the transition $\AP,\vaz\to\AP,\BP$ due to  A ($\AP$)
individuals in each remaining $k-1$ neighbors of the empty vertex ($\vaz$), while
the factor 2 comes from the  transition $\vaz,\BP\to\AP,\BP$ of B ($\BP$)
individuals neighboring the empty vertex. The remaining triplets terms can be
rationalized in the same way.

We now proceed with the pair approximations, $[S,S',S'']=\frac{[S,S'][S',S'']}{[S']}$
in Eqs. \eqref{eq:dA_B} to \eqref{eq:dAB_vaz}, and find that them can be approximated as

\begin{eqnarray}
	\frac{d [\AP,\BP]}{dt} 
	& = & -2(1+{\lambda}/{k})[\AP,\BP]+2\mu [\AP,\ABP] + 2\lambda \frac{k-1}{k} 
	\left(\frac{[\AP,\vaz]^2+[\AP,\vaz][\vaz,\ABP]}{[\vaz]}-
	\frac{[\AP,\BP]^2+[\AP,\BP][\BP,\ABP]}{[\BP]}\right)
	\label{eq:dA_Bpar}
	\end{eqnarray}
 
\begin{eqnarray}
 \frac{d [\AP,\ABP]}{dt}
 & = & -(1+2\mu+\lambda/k)[\AP,\ABP]+ \mu [\ABP,\ABP]  + \frac{\lambda}{k}([\AP,\BP]+[\vaz,\ABP])+ \\ 
 & &  + \lambda \frac{k-1}{k}\left( \frac{[\AP,\vaz]+[\vaz,\ABP]}{[\vaz]} [\vaz,\ABP]
+ \frac{[\AP,\AP]([\AP,\BP]+[\AP,\ABP])+[\AP,\BP]^2-[\AP,\ABP]^2}{[\AP]}\right). \label{eq:dA_ABpar}
 \end{eqnarray}

 \begin{eqnarray}
 \frac{d[\vaz,\vaz]}{dt} 
 & = & 4[\AP,\vaz]-4\lambda \frac{k-1}{k}\frac{[\vaz,\vaz]}{[\vaz]}([\vaz,\AP]+[\vaz,\ABP])\label{eq:dvaz_vazpar}
 \end{eqnarray}

 \begin{eqnarray}
 \frac{d[\AP,\AP]}{dt}  
  &= &  2(-[\AP,\AP]+\mu[\AP,\ABP] +\frac{\lambda}{k}[\AP,\vaz]) +
 2\lambda \frac{k-1}{k}\left(\frac{[\AP,\vaz]+[\vaz,\ABP]}{[\vaz]}[\AP,\vaz] -
  \frac{[\AP,\BP]+[\AP,\ABP]}{[\AP]}[\AP,\AP]\right).\label{eq:dA_Apar}
 \end{eqnarray}

\begin{eqnarray}
 \frac{d [\ABP,\ABP]}{dt}
  & = & -4[\ABP,\ABP] +4\frac{\lambda}{k}[\AP,\ABP]+4\lambda \frac{k-1}{k} \frac{[\ABP,\AP]}{[\AP]} \left(
  [\AP,\ABP]+[\AP,\BP]\right) \label{eq:dAB_ABpar}
 \end{eqnarray}
 
 \begin{eqnarray}
 \frac{d [\AP,\vaz]}{dt} 
 & = & -\left(1 +\frac{\lambda}{k}\right) [\AP,\vaz] + 
\mu [\ABP,\vaz]+[\AP,\AP]+[\AP,\BP]+ \lambda \frac{k-1}{k} \times \nonumber\\
& & \left(\frac{([\AP,\vaz]+[\ABP,\vaz])([\vaz,\vaz]+2[\AP,\vaz]}{[\vaz]}-\frac{[\BP,\AP]+[\ABP,\AP]}{[\AP]} [\AP,\vaz]\right)
\label{eq:dA_vazpar}
 \end{eqnarray}

 \begin{eqnarray}
 \frac{d][\ABP,\vaz]}{dt}
& = & -2(\mu+\lambda/k)[\ABP,\vaz]+2[\ABP,\AP]+2\lambda\frac{k-1}{k}\left(\frac{[\AP,\BP]+ [\ABP,\AP]}{[\AP]}[\AP,\vaz]
-\frac{[\vaz,\AP]+[\vaz,\ABP]}{[\vaz]}[\ABP,\vaz]\right).\label{eq:dAB_vazpar}
 \end{eqnarray}
 
 \end{widetext}


%

\end{document}